# Complexity, Disorder, and Functionality of Nanoscale Materials

Xiaoming Mao,[1,2] Nicholas Kotov[1,3,4]

[1] Center for Complex Particle Systems (COMPASS), University of Michigan, Ann Arbor, MI, 48109, USA;
[2] Department of Physics, University of Michigan, Ann Arbor, MI, 48109, USA;
[3] Department of Chemical Engineering, University of Michigan, Ann Arbor, MI, 48109, USA;
[4] Department of Materials Science, University of Michigan, Ann Arbor, MI, 48109, USA.
Contact authors: maox@umich.edu;   kotov@umich.edu

## Abstract

Nature hosts a wealth of materials showcasing intricate structures intertwining order, disorder, and hierarchy, delivering resilient multifunctionality surpassing perfect crystals or simplistic disordered materials. The engineering of such materials through nanoparticle assembly represents a burgeoning field, poised with potential to yield sustainable material systems rivaling or exceeding biological functionalities. This review delineates the fundamental concept of complexity in the context of nanoscale materials. It examines methodologies for characterizing complexity and functionality, explores pragmatic approaches to create complex nanomaterials, and offers a perspective on their potential applications, guiding the trajectory of future research endeavors.

# 1. Introduction

The need for complexity in materials engineering lies in the pursuit of optimized and robust solutions that address the hard challenges posed by environmental, economic, and social boundary conditions for their design. Sustainable materials must exhibit a difficult-to-achieve balance of properties, such as strength, flexibility, and recyclability, to meet the diverse needs of various applications while minimizing their environmental footprint. It is becoming increasingly clear that traditional crystalline materials (characterized by perfect order) and amorphous materials (characterized by simple uncorrelated disorder) cannot address these needs. Learning from biology, incorporating complexity into material design by combining order, disorder, and hierarchical organization enables the development of innovative structures and compositions that mimic the structural designs of the most advanced and adaptable systems in nature – those of living organisms. To increase the complexity of human-made (or synthetic) materials, whether or not the repeating structural patterns found in living creatures are incorporated – is the best pathway to new materials with enhanced performance while reducing resource consumption. The complexity will also foster their resilience and adaptability of these materials, ensuring that they can withstand dynamic environmental conditions and evolving technological requirements. Embracing complexity is, therefore, pivotal for the development of materials that are foundational for a sustainable future.

The subject of complexity is significant and timely for nanostructured materials because they: (**1**) have an intrinsic ability to self-assemble into complex structures; (**2**) require deliberate design of intricate architectures to best ensure their properties like conductivity, mechanical strength, and catalytic activity; and (**3**) provide the unique opportunity to observe the emergence of complexity over multiple scales due to high electronic contrast and microscopic observations of nanoparticle (NP) dynamics. As the design of complex self-assembled nanostructures continues to advance to benefit energy, environmental, water, biomedical, information and other technologies, a quantitative approach to complexity engineering of nanostructures becomes the forefront of materials science and technology.

The objectives of this perspective are threefold: (**1**) to define the notion of complexity in the context of materials; (**2**) to establish and validate methodology(ies) of its quantification; and (**3**) to provide practical pathways for further scientific studies of complex particles, composites, assemblies, and biomaterials and their technical and biomedical realizations.

## 2. Historic Perspective: What is Complexity?

The subject of complexity has many faces. It represents a ubiquitous yet vague notion that does not have a universal agreed-upon definition based on the physical description of matter. The earliest thought about complexity traces back to Aristotle, who defined complexity as "*The whole is more than the sum of its parts*" (Metaphysics, Vol. VII). Note that this is a function-based conceptualization of complexity, which can theoretically guide the construction of many complex systems and materials.

The first modern attempts to transition from colloquial to formal definitions of complexity are associated with the development of programming and information theory. The works of Kolmogorov,[1] Solomonoff,[2] and Chaitin[3] in the early 1960s described *complexity* as the minimal algorithm required to fully reproduce an object and its information content. Although all three of these scientists worked on this subject independently, this concept is often referred to as *Kolmogorov complexity;* we will refer to it as *algorithmic and information complexity* or *AIC*. The practical realization of AIC is a minimal set of symbols that fully encodes the composition and structure of an object. The AIC of a macroscale material reflects the minimal number of structural rules and parameters capable of describing its current organization and future dynamics.

Repeatable patterns found in crystals, dramatically simplify the algorithms (rules) describing their structural organization, and thus AIC. At the same time, randomness of the atomic structures of the materials increases AIC.

The relationship between Shannon information, algorithm design, and complexity was further explored in the 1980s in the publications by S. Wolfram in 1984,[4] and P. Grassberger in 1986.[5] The theories of cellular automata, chaos, and fractals brought about a different notion of complexity in the studies by A. B. Çambel in 1992,[6] S. A. Kauffman in 1993,[7] M. Gell-Mann in 1994,[8] Lopez-Ruiz et al. in 1995,[9] and J. Holland in 1995.[10] Using information theory and AIC as stepping stones, their papers and books promulgated the concept of complexity as a *combination of order and disorder* (COD). Perfect periodicity and long-range order reduce the number of rules governing the organization of an object. Thus, crystalline materials, in the context of both COD and AIC, are much less complex than glassy materials. However, Çambel, Kauffman, and Gell-Mann treated randomness as one of the algorithmic rules governing the organization of a system. A fully stochastic system, therefore, becomes *noncomplex* and the information content of two random distributions of atoms becomes identical despite the difference in their coordinates, which is in full agreement with physical realities of large atomic structures. An increase of a system's randomness will therefore constitute a decrease in COD as opposed to an increase of AIC. Another difference between these two concepts is that COD is no longer bound to any computational algorithms.

The change in physical interpretation of complexity enabled these scientists and engineers to explore the relations between complexity and "big" subjects, such as the emergence of life and evolution of the universe. Elegant, instructive, and quantitative insights into the relationship between complexity and thermodynamics were also made. The direct connection between the complex multifractal systems of particles and their entropy and enthalpy were made by Stanley and Meakin in 1988[11] The relationships between statistical thermodynamics and complexity were expanded by Lopez-Ruiz et al. in 1995[9] who showed the rise of complexity as the system moves away from equilibrium. The physics of complexity was further elaborated by P. Bak[12], C. Tang, and K. Wiesenfeld who promulgated the idea of "self-organized criticality" (SOC). This phenomenon describes a dynamic macroscopic system that spontaneously evolves towards criticality and scale invariance, which is considered as a mechanism for the emergence of complexity in chaotic and noisy environments.

The interest in complexity continued between 1990 and 2000 and expanded to the structural complexity found chemical and biological matter. These studies included brain and neuronal networks,[13] biological organisms,[14] chemical systems,[15,16,17] and nanoscale structures.[18] Particularly significant was an idea that complexity describes the structural sophistication of matter based on functionality rather than information content,[19] which was also advocated by Kauffman[7] and Gell-Mann[19] Many of these advances in the theory of complexity were stimulated by organization of the Santa Fe Institute - a private research organization focused on understanding complex systems that Gell-Mann helped to establish.

## 3. Complexity and Functionality

### *3.1. Functionality-based complexity*

Expanding the ideas of Aristotle, Grassberger, Holland, and Gell-Mann about an *effective* and *function-based complexity*, let us consider the reasons for complexity as a design metric for materials. The best examples of amazingly complex and universally useful chemical structures excelling in both performance and longevity are biological composites. They can be found in all parts of the biosphere and are typically made from renewable earth-abundant nanoscale components, which makes a particularly good case for complexity in the context of sustainability. High-performance biological composites are exemplified by seashell nacre,[20] tooth enamel,[21] articular cartilage,[22] tree roots,[23] (**Figure 1**) and animal bones[24] (**Figure 2**). Their complexity manifests in sophisticated geometrical patterns, multiplicity of the components, abundance of interfaces, and multiple scales of organization. All these structural aspects are needed for them to perform their intended duties while reducing the energy burden on the organism and maximizing access to the material's components. An iconic example of complex biomaterials are bones with organizational motifs spanning multiple scales from $10^{-10}$ m to $10^1$ m with all of them being essential for their combination of functionalities that include lightweight load-bearing, stem cell differentiation, immune response modulation, and others (**Figure 2**).[25,26] At the atomic and nanoscale levels, bones are based on NPs with distinct crystallinity[27] The NP shape is asymmetric, and they are exceptionally polydisperse[28] The mesoscale organization of the NPs in different projections includes mutual alignment and complex stochastic patterns with distinct helicity (**Figure 2 C,D**).[28] The microscale and submillimeter scale organization display characteristic diameter of pores, asymmetry of niches, and thickness of pore walls.[29] What emerges from an overview of these and other biomaterials as well as their human-made replicas is that the organizational pattern and constitutive building blocks can be different, but the combination of order and disorder is universal.

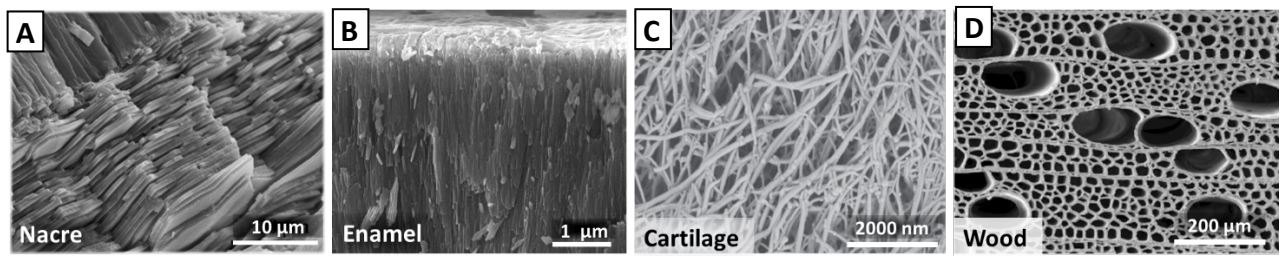

**Figure 1.** Complex biomaterials. Scanning electron microscopy images of (**A**) nacre;[20] (**B**) tooth enamel;[21] (**C**) articular cartilage;[22] and (**D**) tree roots (poplar).[23]

Page 6

The omnipresence of the correlated disorder and its direct relevance to the complexity and performance of the materials can be seen from different standpoints and multiple scales of organization. As depicted in **Figures 1** and **2**, the architecture of load-bearing materials exhibits several universal structural traits: (**1**) Atomic periodicity of NPs is paired with amorphous layer at their interfaces[30] (**2**) Mineral components display consistent anisotropy, but they are never the same size or shape; (**3**) Mesoscale alignment of nanorods or nanofibers is distinct and common, but it is also variable and imperfect even for tissue segments with similar functional requirements; and (**4**) Microscale and millimeter scale patterns have non-random pore dimensions, but their percolating patterns and wall connectivity are seemingly disordered. Identical observations can be extended to all high-performance biomaterials.

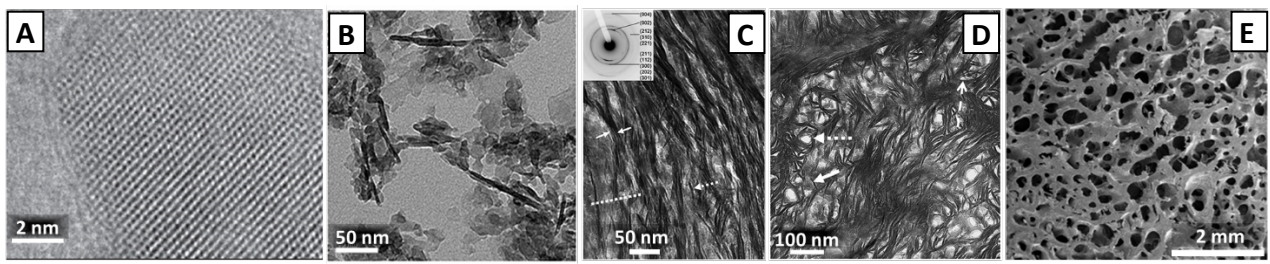

**Figure 2.** Examples of hierarchical organization of complex biomaterials. Transmission (**A-D**) and scanning (**E**) electron microscopy images of bone structure at different scales: (**A, B**) constitutive hydroxyapatite nanoparticles; [25, 26] (**C, D**) mesoscale assemblies of nanoparticles in different projections; [26] and (**D**) microscale organization of hydroxyapatite nanocomposite.[29]

The key reason for universality of these structural traits is that the combination of order and disorder enables hierarchical organization, which would be otherwise impossible in a perfectly ordered atomic or molecular, nanoscale, or microscale crystal. Disorder is precisely what is needed to accommodate the scale-dependent 'switch' from one organizational pattern to another required for survival-critical physical properties determined organization of matter at different scales, for example, hardness and toughness, or light scattering and nutrient transport.

Note that the intricate multiscale organization have, however, enthalpic and entropic penalties. Noncomplex materials, for instance, perfectly crystalline minerals or perfectly disordered glasses, are more advantageous with respect to the free energy costs and component availability. Nevertheless, the materials with high complexity are the ones that make a difference between existence and non-existence for living organisms. They support life because they excel not in only one task but in multiple ones at the same time. Nacre, enamel, bone, cartilage, etc. require complexity because they must display high strength, high toughness, high ionic selectivity, high flux of nutrients, low weight longevity, and selectivity in cellular adhesion – all at the same time. Monolithic blocks of calcium fluorophosphate, collagen, or their mixture exhibiting complete disorder or perfect order (crystallinity) will not provide the hardness, strength, and longevity of enamel.[21] This property set requires it to be made from hydroxyapatite nanorods oriented along the dominant direction and interlaced with protein cushions.

Besides examples from nature, there are numerous materials, and more generally, materials *systems*, that are neither periodic nor random, which are essential for their performance, robustness, and manufacturing. They span a vast range of technologies from neuromorphic computers to desalination membranes.

Considering the prior studies of complexity in the context of 'big questions', a parallel can be made with other complex systems evolving under multiple boundary conditions modelled by Kauffman[31] and Holland[10]  Similarly, complexity in biomaterials emerges because they must perform multiple tasks and conform to multiple functional requirements that would otherwise be impossible to fulfil.

In the context of materials *complexity* can, therefore, be defined *as purposeful performance-oriented structural organization of matter, combining order and disorder*. This definition is applicable to biological and non-biological materials, static or dynamic matter, and open or closed systems. It also incorporates atomic, nanoscale, mesoscale, microscale, and macroscale structures relevant to understanding the interactive and malleable hierarchy in living and technological systems.[32] The combination of order and disorder makes possible combining different structural motifs at multiple scales, which de facto incorporates the notion of "*The whole is more than the sum of its parts*," where at each scale, bringing together different parts of the material gives rise to new complex properties, maximizing functionality.

*3.2. Quantification of Complexity*

The approach to complexity as a parameter relating materials organization and performance can be described as a generic "Goldilocks" curve (**Figure 3**). Based on the necessity of materials structure to satisfy multiple requirements giving rise to multiple properties, the complexity and performance reaches a maximum somewhere between order (i.e., perfect continuous crystallinity) and disorder (i.e., uncorrelated isotropic randomness).[8] The earliest version of such a curve was found in Huberman and Hoggs' 1986 publication.[33] Similar curves also appeared in subsequent publications by Lopez-Ruiz et al. in 1995,[9] B. Edmonds in 1995,[34] and Tononi at al. in 1998[13] Analogous conceptual dependencies between complexity and regularity also appear in very different subjects, such as art.[35,36]

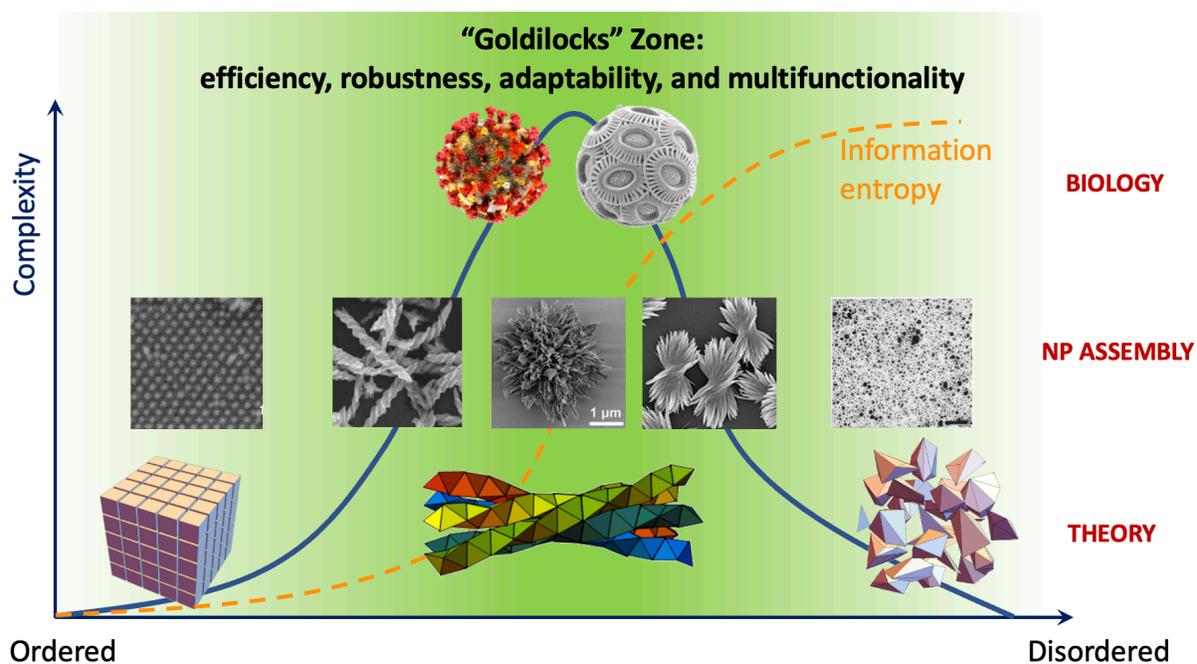

**Figure 3.** Effective complexity peaks between ordered and disordered limits, defining the "Goldilocks" zone where complex biological structures arise with outstanding features (top row). Simple measures of information entropy monotonically increase with disorder. However the effective, functionality-based complexity peaks in the middle. Functional complexity of NP self-assembly (images, middle row) can be studied using theoretical approaches such as geometrically frustrated polyhedral assembly (bottom row).

We note that the axis of complexity in all the Goldilocks curves proposed in this field lack a quantitative assessment of complexity. This vagueness represents a major roadblock for future studies and, theoretically, both AIC and COD interpretations of complexity can be applied to bring these curves closer to reality.

Multiple approaches to quantify complexity have been proposed in the past.[37] The string of symbols with minimal possible length for an ideal Turing computer to reproduce an object was a logical measure of AIC. One obvious problem here is that AIC is dependent on a programing language. A bigger problem, however, is that AIC is incomputable for any realistic object,[38] which becomes particularly obvious for dynamic macromolecular structures relevant for chemistry or biology, as exemplified by a nanoscale particle or a polymer chain surrounded by water molecules. On the positive side, AIC can be related to entropy and entropy-related measure(s) of complexity were proposed by Grassberger[39], Rajaram, and Castellani.[40] AIC inspired measures of complexity based on statistics integrated with thermodynamics were put forward by Crutchfield and Young in 1989[41] as well as by Lopez-Ruiz, Mancini, and Calbet in 1995.[9]

Notwithstanding the significance of entropy and information content to practicality, the examples of biomaterials shown in **Figures 1-2** indicate that measures of COD would be a better choice for materials engineering than AIC. Measures of COD based on geometrical parameters of the system were proposed by Tononi et al. in 1994,[13,42] Ay et al. in 2011[43] and Wolf et al. in 2018[14] Notably, all these approaches were inspired by biological systems. Cumulatively, they can be described as different methods to divide the actual physical system (Tononi, Ay) or its photographic image (Wolf, Katsnelson, Koonin) into parts followed by testing whether the complete system (object, image, etc.) has more information than a sum of its parts. In this regard, one system is higher in complexity when its generalized functionality is higher than that of another system. This approach and its different interpretations capture the idea of and can theoretically lead to a quantitative Goldilocks curve.

*3.2. Complexity of Nanoscale Assemblies*

Self-assembled nanostructures can serve as a convenient model to further our understanding of how complexity and functionalities are related. Varied degrees of order and disorder are typical for NPs and other nanoscale structures. The latter includes a range of structures from nearly perfect colloidal crystals made from NPs with simple singular functionalities to imperfect but multifunctional assemblies from polydispersed NPs and to random agglomerates with limited functionalities. Additionally, the structures of inorganic nanoscale assemblies can be evaluated by various microscopy methods because of their higher contrast compared to of inorganic materials. Finally, the variety of nanoscale materials in biology and technology is enormous. The creation of a universal approach to their design that incorporates multiple requirements would significantly simplify the job of materials scientists.

Prior studies of complexity were not adapted to nanoscale structures in part because a methodology for the structural description of nanoscale systems that combined both order and disorder had not yet been defined.[14,43] In one recent case, the architecture of nanoassemblies were analyzed using the toolbox of Graph Theory (GT).[44] In this approach, NPs are represented as nodes, while the structural connectivity between them (both static and dynamic) is represented by edges (**Figure 4**). This general description can be universally applied across all material platforms with nanoscale organization. Importantly, the different shapes of the NPs, despite their variability in size, can be represented using simple GT formalisms. For example, nanospheroids, nanorods, and nanoplatelets can be described as fully connected $K_1$, $K_2$, and $K_3$ graphs. Chiral 3D structures with mirror asymmetric shapes can be represented by $K_5$ graphs. All these representations of the constituent nanoscale building blocks correspond to the graphs of minimal complexity. Simplification of graph representation of the nanoscale components is needed to have consider contribution from unifying connectivity patterns between the building blocks. This approach allows for the calculation of the values of complexity that are most applicable to both fundamental studies and practical implementations.

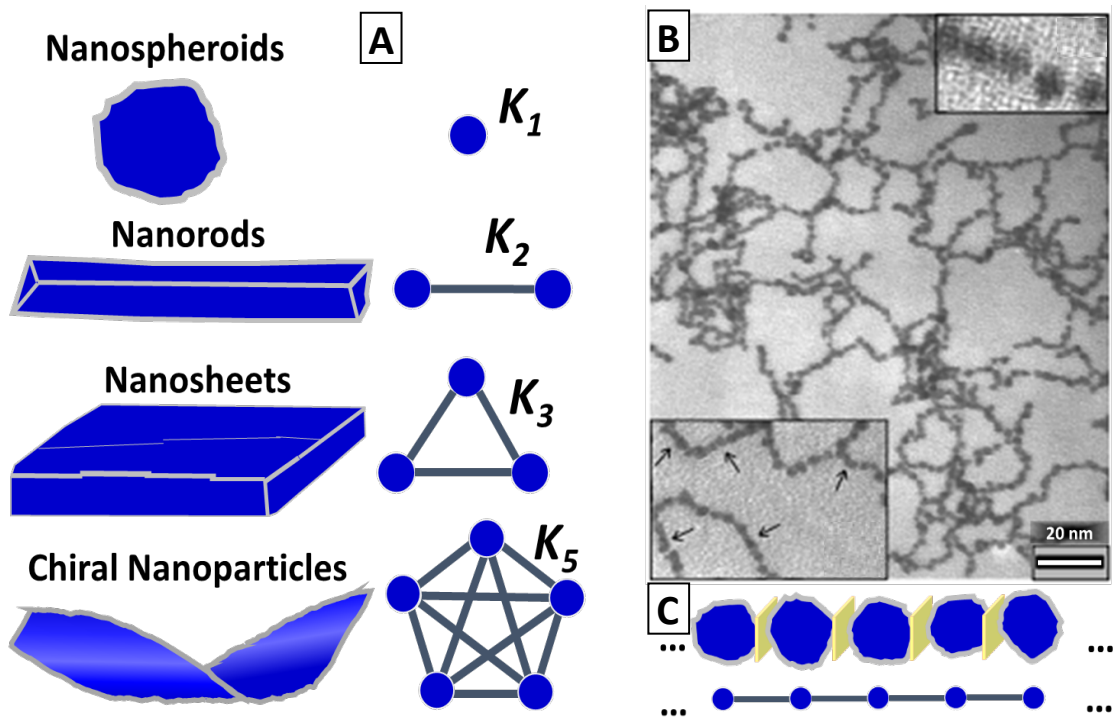

**Figure 4.** (**A**) Basic concept of GT description of NP assemblies. NPs with more complex geometry are represented as $K_n$ complete graphs with increasing number of nodes depending on the shape and symmetry elements.[5] (**B**) TEM image of chains of CdTe NPs forming a network and (**C**) their GT representation based on $K_n$ formalism.

GT models of nanoscale structures and nanoassemblies can be extracted from their TEM and SEM images (**Figure 4B, C**).[45] Most certainly, high-resolution 3D reconstructions of nanoscale materials are preferred and currently available in many research centers and universities. Note that graph descriptions of nanoscale structures can be applied to both size-limited nanoscale assemblies from NPs, nanorods and nanoplatelets, such as supraparticles, as well as to (semi)infinite materials, such as composites, gels, etc. The GT approach can be extended to continuous materials, such as nacre, enamel, and bone.[44] For long-range structural patterns with high stochasticity, descriptions of nanoassemblies can be made in terms of random graphs and networks.[46–48]

For a complexity measure to serve as design parameter for materials with specific property sets it must reflect the hierarchical organization of the materials system. The combination of order and disorder can be used as a tool to attain the functionalities with different scale-dependences and those controlled by parameters in molecular, nanometer, micrometer, and other scales. The GT toolbox can describe the hierarchical organization of matter as graphs also. For example, the atomic scale structure of NPs, their interface with polymers, and the particle chains that form the polymer can be captured by GT. While possible and practical, such descriptions have not been yet realized and will require the introduction of additional variables such as nodal and edge weights that can establish the specific scale and geometry.

Once a representative graph of a nanostructure, or more generally, a material, is created, its complexity can be calculated. However, there is no universally accepted measure of complexity (Section 3.2). Additionally, measures developed in the classical studies of complexity have not yet been applied to materials – nanoscale or otherwise. Several complexity measures could potentially be applicable to nanostructures. Among them are fractal exponents that describe self-similarity across the scales calculated from the GT representations.[49–53] The quantitative parameters of resulting fractal exponents, that is, the pre-factor and fractal dimension, $D$, are often associated with complexity[54] but neither pre-factor nor fractal dimension are measures of AIC or COD. The concept of singular fractality could be suitable for the description of hierarchical organization of materials when the patterns are repeatable at different scales. However, this is not true for the actual materials (**Figure 1,2**). One of the reasons for combining order and disorder is the need to change structural patterns for different scales to accommodate realization of different properties.

Organizational patterns with multiple structural motifs can sometimes be described by a single fractal exponent, such as perfectly ordered fractals represented by Sierpinski triangles, and random isotropic fractals represented by diffusion-limited agglomeration structures. In a more general case, one can use a spectrum of fractal dimensions treated as scale- and location-dependent variables. These methods to describe repeatable patterns are referred to as multifractals and multifractal spectra.[55,56] They can be a promising step toward structure-property relations for complex materials. While applications of multifractality to materials science are being developed, dependences of macroscale properties on multifractality spectrum extracted from GT representations can be a promising new path for research.[57–59]

Another approach is the development of specialized indexes, such as the *complexity index* (*CI*).[44,60] The calculation of *CI* is based on GT representations of nanoscale structures for which the complex organization of the material combining order and disorder are described based on the $K_n$ formalism for nanoscale structures. The use of $K_n$ enables one to abstract the structural imperfections and variability of the components that make up the material. *CI* for nanostructures

Page 14

is designed to identify the repeatable patterns of their organization. For example, the structure of NP chain can be represented by a simple graph formula in **Figure 4C**. The similarity with common chemical formulae is not accidental because they are indeed atomic graphs. The structure of more complex nanostructured materials, such as supraparticles from achiral nanosheets and nacre can be represented by the GT formulae in **Figure 4A**, where the platelets or platelet-like components are represented by $K_3$ graphs. Two key differences with atomic graphs are that (1) the NPs can have variable shapes including those that are atypical for atoms (**Figure 4A**); (2) equality of sizes is not required and, unlike atoms, individual NPs are expected to have imperfections.

Their repeatable organizational pattern of nacre or nacre-like man-made composites, that is, the stacking of nanoplatelets with organic layers in between, is uniformly described by a graph where $K_3$ segments are interconnected by additional edges (**Figure 5A**). Alternatively, the random distribution of nanoplatelets in supraparticles with repeatable nonrandom size is described by the loop with $K_3$ segment inside (**Figure 5B**). GT representations of other complex particles both man-made (**Figure 5C-D**) and natural (**Figure 5E**) can also be built. *CI* is then calculated as the limited sum of $(½)^n$ progression for the total number of edges connecting each type of node to it nearest neighbors, next-nearest-neighbors, second-next-nearest neighbors, etc. If there are more than one type of node, the *CI* for each of them are added together. Calculations of *CI* were first used when the materials with a combination of order and disorder, for instance, nacre with polydispersed platelets with stochastic yet nonrandom shapes were considered.[44] Similarly, complex hedgehog supraparticles with randomly organized cores and a non-randomly organized halo of spikes were successfully described by GT formalism where random portions of the particle with yen nonrandom diameter was described by loops.

*CI* can serve as a measure of COD complexity. For complex particles, the rise in *CI* can be directly associated with their multifunctionality, and potentially construct a Goldilocks curve. The rise in functionality is characterized by the multitude of properties controllable by organization at these scales. The addition of atomic scale that describes the nanoparticles, nanorods, nanosheets, and the heterogeneity over the volume of the supraparticles, will increase the degrees of freedom, the number of nodes in GT representations, and thus, the measure of complexity.

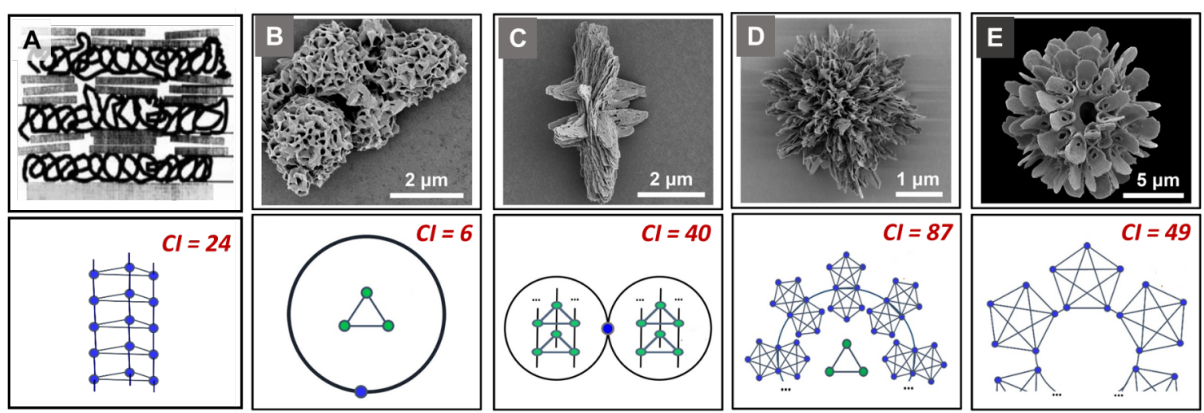

**Figure 5.** Structural images (top) and GT representation (bottom) for various nanoassemblies (**A**) nacre-like composite represented stacks of graphite oxide nanoplatelets; (**B**) supraparticles from achiral Au-S nanosheets; (**C**) kayak particle from achiral Au-S nanosheets; (**D**) coccolith-like particles from chiral Au-S nanosheets, (**E**) skeleton of the algae *Syracosphaera Anthos* containing chiral microsheets.

Concomitantly, this will also increase the number of properties that can be combined for a particle and the range of parameters that one property can be tuned independently of others. A recent proposal in this workspace was to perform real space renormalization on images and calculate the distances (i.e., dissimilarities) between images across consecutive steps of renormalization, and use the sum of these distances across the scales to define a multiscale structural complexity.[61] This definition captures the variation of the structure as one zooms into smaller and smaller scales, which aligns with our definition of complexity in nanomaterials in terms of structural variation across scales.

Page 16

## 4. Pathways to Complexity

How can one create a complex material? Answers to this question hold the key to technologically friendly pathways to scalable complex materials that exhibit a desired set of properties. Although many views on complexity have been proposed and many methods to predictively describe complex systems have been postulated, the actual pathways to *real materials from real components* that exhibit intended complex behavior and an intended set of properties, have been much less explored. As often is the case with such difficult problems, some lessons can be learned from examining complex materials in nature. Most of them are nanostructured, self-assembled, and chiral. In this section we review several models that connect these characteristics with COD complexity.

*Chirality:* Chirality, or mirror asymmetry, arises from the spatial arrangement and chemical distinction of atoms, molecules, nanoparticles, supraparticles, etc. This geometric characteristic and associated parameters introduce a uniquely practical dimension to the topic of material complexity. In chemistry and biology, it is widely accepted that chiral components self-assemble into intricate and sophisticated structures, which often exhibit properties that are very distinct from their initial components. Chiral molecules and particles can also be synergistically combined with non-chiral counterparts. The arrangement of chiral entities, such as molecules or NPs, can lead to the emergence of materials with complex helical or twisted structures.[62]

Understanding the relationship between the complexity of desirable materials and the chirality of their components provides a path to synthesize materials at large scale while carefully tuning their mechanical, optical, electronic, and biological properties. Although publications of empirical observations of complex materials with chiral molecules and particles are copious, the physicochemical mechanisms leading to their geometrical complexity are still puzzling. This is particularly true for nonuniformly sized nanoscale components that are essential for the scalability of certain processes. A direct inquiry into the dependence of chirality and complexity was made using a model system of supraparticles (SPs) from polydispersed gold thiolate NPs with the shape of nanoplatelets ($K_3$). Because the surfaces of the NPs were coated with *L*- or *D*-cysteine (Cys) ligands, the flexible nanoplatelets twisted to become chiral. The resulting SPs then required $K_5$ representation in the GT models. There was a clearly visible increase in the complexity of the self-assembled SPs as chirality increased as a result of changing the *L/D*-ratio from 1 to 10 (**Figure 6**). The calculation of *CI* values for SPs of different morphologies made at various temperatures and *L/D*-ratios demonstrated highly non-linear growth of complexity as the chirality of the system increases.

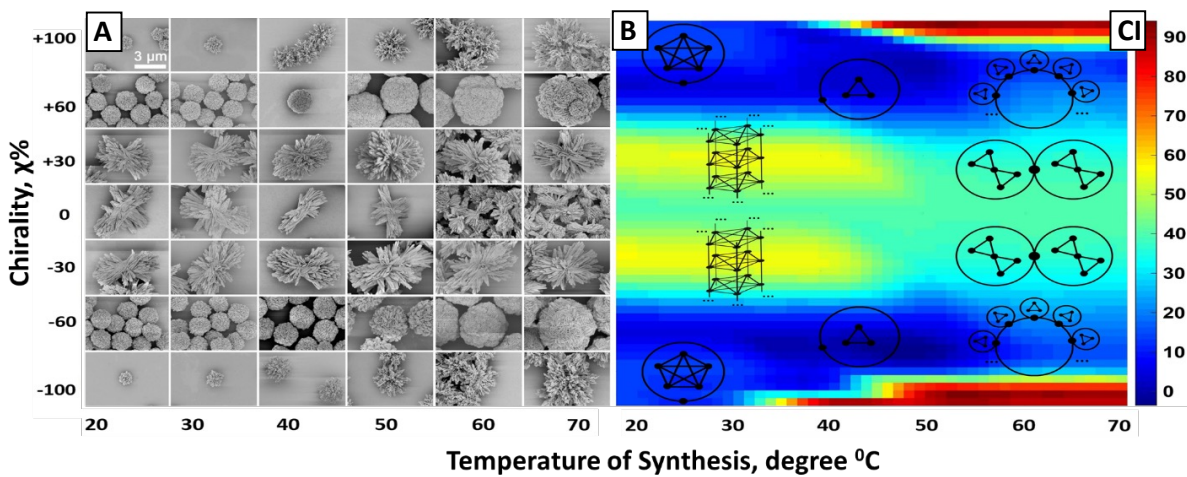

**Figure 6.** (**A**) Phase diagram observed for complex particles from chiral Au-S nanoplatelets. The blue and red phases on the top and bottom correspond to coccolith-like particles with high complexity in **Fig. 1b**. (**B**) Mapping of *CI v*alues on phase diagram in (**A**). The highest *EC* values are observed for the part of the phase diagram where a triple point specific to critical states is observed. χ is the enantiomeric excess of chiral amino acids coating the Au-S platelets. $t_n$ is the temperature of the reaction.

This dependence can be understood in the context of competitive restrictions on the short- and long-range self-assembly patterns that emerge for systems with different chirality. Electrostatic repulsion between the components engenders the nearly universal restrictions that impose limits on particle size, which is common for SPs, vesicles, micelles, inkjet droplets, exosomes, intracellular compartments, etc. The complex nanoscale chiral architectures emerge only when restrictions are multiple, anisotropic, and competitive. What is essential is that energy gains and penalties associated with these restrictions on the self-assembly of building blocks must be comparable.

When none of the competing interactions and restrictions dominate, NP self-assembly acquires complex architectures to negotiate these conflicting requirements. For chiral NPs, the anisotropic restrictions associated with electrostatic repulsion are intertwined with those from hydrogen bonding, hydrophobic attraction, meta-ion coordination, and elastic deformation. The first order assessment of characteristic energies of these interactions when competitive restrictions result in complex architectures is 50-60 kJ/mol.[44] Since the shape, charge, elasticity, hydrogen bond density, and composition of NPs change with *L/D* ratio and temperature, the characteristic energies also change. Thus, chirality provides the essential structural element at angstrom and nanometer scales for multiple restrictions to become competitive. Also important, chirality is a scaleless geometric property and, therefore, under favorable conditions it can be transferred from scale to scale and to higher scales, thereby ensuring competitive restrictions and complexity evolution as the size of the objects become larger.

The practicality of the chirality pathway to complexity can be highlighted by the fact that the polydispersity of NPs does not impede the formation of complex architectures. For Au-Cys NPs with extremely high polydispersity, the complexity of the resulting SPs is considerably higher than those formed from other NPs. Polydispersity may actually *increase* the complexity of the supraparticles, because the presence of smaller and larger NPs with stronger and weaker twists enables the growing assembly to select site-suitable NPs from the mixture.[63] The uniqueness and tunability of the optical, chemical, colloidal, and potentially biological properties of these complex SPs suggest their broad application in asymmetric catalysis and polarization-based optoelectronics. The strong effect of chirality enables the scalable preparation and controllable preparation of complex particles in a sustainable manner and helps us to understand the origin of the astounding diversity and sophistication of biological nanocomposites (**Figures 1,2**). Further directions in this area may also include evaluation of the relations between COD complexity and chirality measures, exemplified by Hausdorff distance, Osipov-Pickup-Dunmur index, Continuous Chirality Measure, and Helicity measure. Also note that Harris, Kamien, and Lubensky showed that, instead of a simple "handedness" parameter, an infinite hierarchy of chiral moments can be utilized as chirality measures. [64]

*Spatiotemporal chaos:* The term chaos refers to persistent random behavior resulting from deterministic equations with initial conditions. The logistic map is perhaps the simplest example of chaos where complexity emerges from one simple iterative mapping of $x_{n+1} = r\, x_n(1 - x_n)$ for $r > r_c \approx 3.57$ through the onset of chaos, where orbits become irregular and small variations in the initial conditions changes drastically over time (**Figure 7A**).[65] In this case, AIC complexity is small because the pattern is generated from a simple equation. COD complexity is, nevertheless,

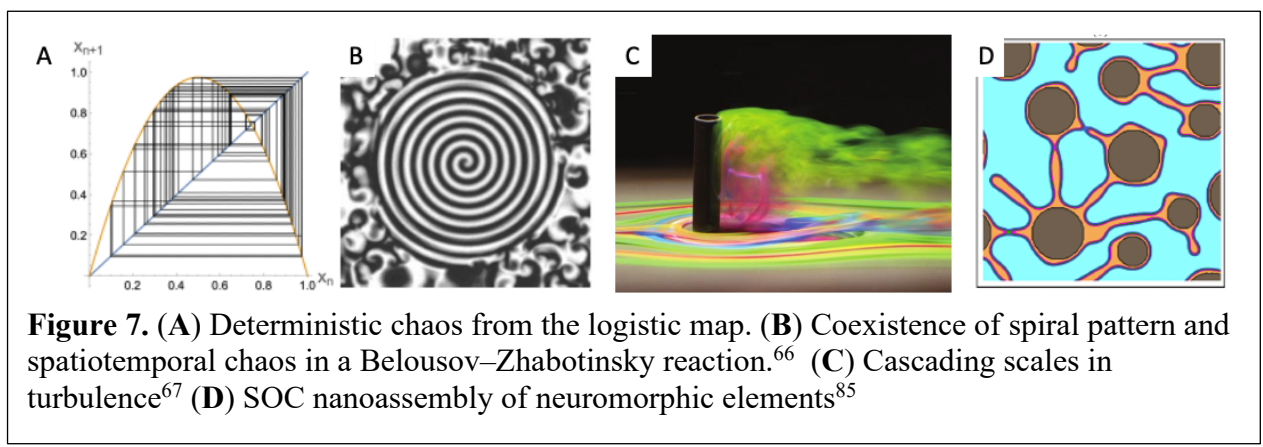

**Figure 7.** (**A**) Deterministic chaos from the logistic map. (**B**) Coexistence of spiral pattern and spatiotemporal chaos in a Belousov–Zhabotinsky reaction.[66] (**C**) Cascading scales in turbulence[67] (**D**) SOC nanoassembly of neuromorphic elements[85]

high with cascading time scales, where temporal correlation function vanishes at long times, indicating that the information of the initial conditions is "forgotten".

The more generalized concept, *spatiotemporal chaos*, describes extended systems involving many interacting degrees of freedom, and finite correlation in both time and space controlled by the time elapsed from the onset of chaos. Having "finite correlation in both time and space" is common in systems with stochasticity, such as statistical mechanics, but spatiotemporal chaos refers to how this phenomenon comes from nonlinearity, even when the system is fully deterministic. This concept is directly relevant for physical, chemical, and biological systems where complex spatial patterns arise from simple nonlinear dynamics rules.[66] Other well-known examples include turbulence[67] and spatial temporal phases in chemical pattern formation systems[68] **(Figure 7B,C)**.

The connections between spatiotemporal chaos and complex biosystems in nature have been extensively discussed in prior studies. There is a high degree of certainty that spatiotemporal chaos can be a realistic pathway to complex materials. Similar to many cases of complex materials, this research direction remains largely unexplored for nanostructures and other materials.

<u>Geometric frustration:</u> A new route to complex structures that has been identified recently is geometric frustration between the building blocks, which can take the form of frustrated attraction and repulsion **(Figure 8A)**, or unfitting geometric shapes. Inability of the geometrical shapes to perfectly tile surfaces and fill 3D volumes without gaps or overlaps, defines the commonly accepted understanding of geometric frustration, and leads to complex assemblies. Geometric frustration has far-reaching influences in statistical mechanics and condensed matter physics in many contexts from glass transitions, spin glasses, ergodicity breaking, frustrated magnetism, and quantum spin liquid states, giving rise to an abundance of complex phenomena. It has also been considered as a plausible component in conjunction with SOC of biological complexity,[14] where geometric frustrations and other restrictions lead to both compartmentalization and rugged fitness landscapes with non-ergodic evolution.

Geometric frustration may be caused by chirality **(Figure 8B,C)**. A good example is the assembly of twisted bundles described by Grason and Bruinsma in 2007 **(Figure 8C)**, where the relative twist between neighboring fibers in a bundle assembly leads to the overall helicity.[69] The authors showed that accumulation of stress and self-limited growth are key outcomes of geometric frustration in self-assembly. More generally, geometric frustration can be viewed in a broad class of "shape incompatibility" problems, which show up in many different forms, from ill-fitting polygons **(Figure 8D)**[70] to twisting cubes[71] and frustrated tubes.[72] Among these challenges, a new route to mathematically formulate and solve the problem was to find the non-Euclidean space where the shapes perfectly tessellate. This theory has been applied to tetrahedra, for which the ideal tessellation is the 600-cell characterized by positively curved 3D space (the 3-sphere, or S-3), which converges to assembled helicoids, in agreement with experiments **(Figure 8E)**.[73] The generality of the theory was illustrated by its application to icosahedra, which tessellates negatively curved space (the hyperbolic 3-space, or H-3), as well as rationalizing Euclidean space self-assembly.[74]

*Far-from-equilibrium assembly:* Kinetic processes far from equilibrium (FFE) generate particle-based structures combining order and disorder as exemplified by snowflakes and fractal nanoassemblies.[75] These are distinct from aggregates formed during spatiotemporal chaos where nonlinear interactions lead to complexity. Fast growth and self-regulation via feedback, demonstrated by exhaustion of the particle supplies, are the main mechanisms for complex assembly in FFE processes. A variety of experimental systems of this sort have been explored, although a unified theoretical framework is still lacking, due to the challenging nature of non-equilibrium statistical mechanics.

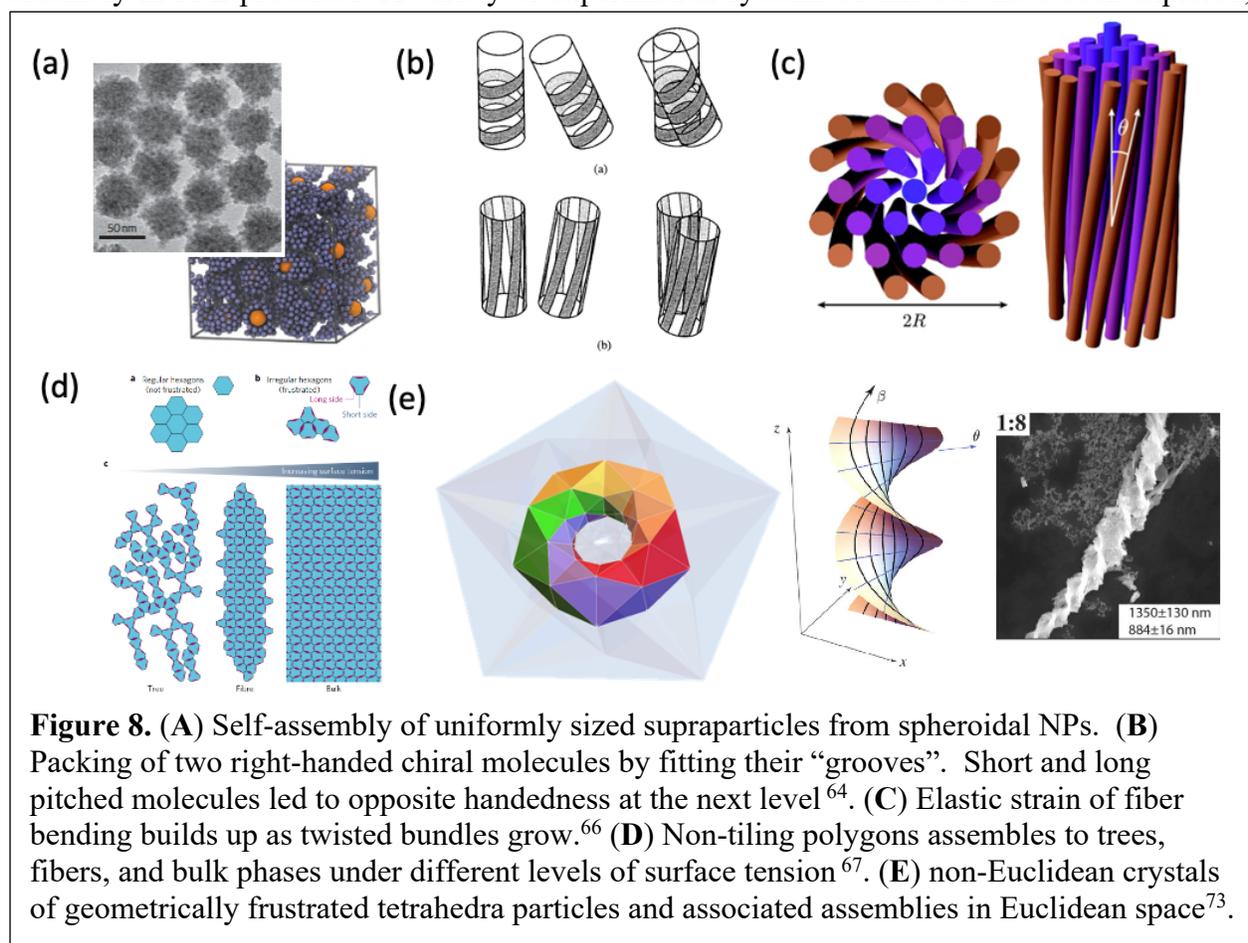

**Figure 8.** (**A**) Self-assembly of uniformly sized supraparticles from spheroidal NPs. (**B**) Packing of two right-handed chiral molecules by fitting their "grooves". Short and long pitched molecules led to opposite handedness at the next level [64]. (**C**) Elastic strain of fiber bending builds up as twisted bundles grow.[66] (**D**) Non-tiling polygons assembles to trees, fibers, and bulk phases under different levels of surface tension [67]. (**E**) non-Euclidean crystals of geometrically frustrated tetrahedra particles and associated assemblies in Euclidean space[73].

A well-known case of FFE assembly is the dendritic growth of crystals **(Figure 9A)**, originating from branching instabilities at growth interfaces, with snowflakes being the simple example. The disordered version of such branching assembly, diffusion limited aggregation (DLA, **Figure 9B**), have been observed in a wide range of systems across ionic, atomic, polymeric, nanoparticle, and colloidal particle scales[76,77]. An important case of such growth is the chemical conversion of ions into dendrites[78](representing a safety problem in batteries and other energy storage devices.[79]) [76,77] Such branching growth under FFE conditions leads to interesting complex structures such as multifractals.

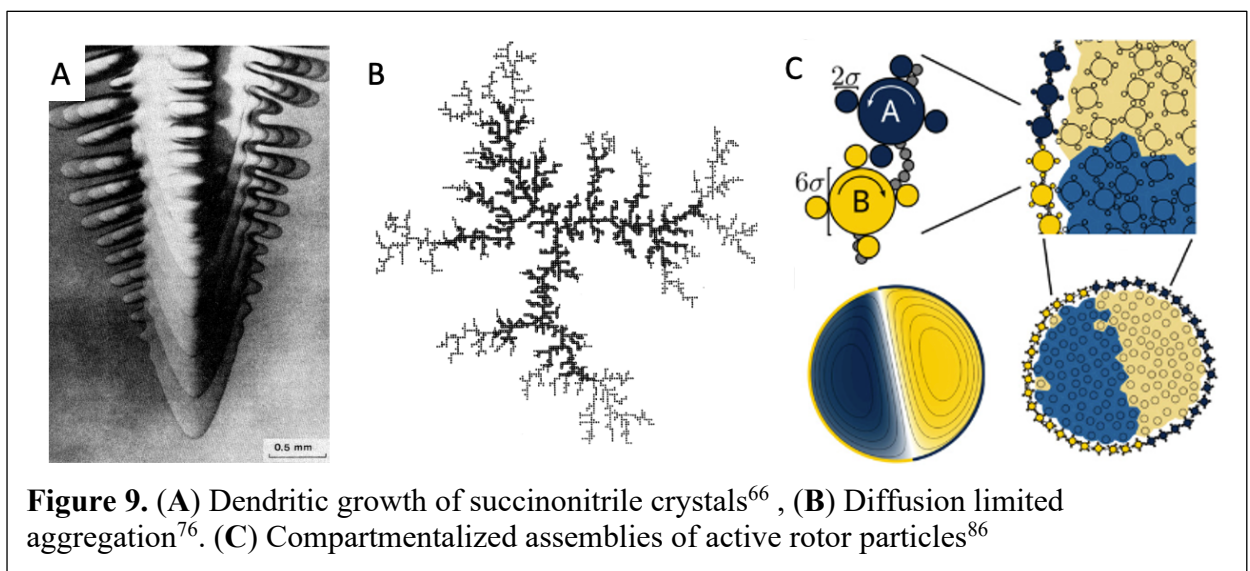

**Figure 9.** (**A**) Dendritic growth of succinonitrile crystals[66], (**B**) Diffusion limited aggregation[76]. (**C**) Compartmentalized assemblies of active rotor particles[86]

More recently, influenced by the blossoming field of active matter,[80,81] using active driving at the particle level to control self-assembly has become a fruitful new research direction and led to the formation of interesting complex structures Directional motion and rotation of the self-assembly building blocks have led to the discovery of a variety of active assemblies [82]Among their unique properties motility-induced phase separation was observed **(Figure 9C)**.[83]

*Self-organized criticality:* An interesting and potential general pathway to complex systems based on molecules and NPs is SOC. SOC links nonlinearity with critical scaling, fractality, and robust complex behaviors, where nonlinear dissipative systems spontaneous evolve to criticality, showing long-ranged spatiotemporal correlations without fine tuning. Frustrated assemblies, for instance, SPs, can be treated as quasi-equilibrium states,[6] which enables the use of extensive statistical thermodynamics to select conditions for their formation. A strong indication that SOC pathway to complexity is practical are the recent data on spontaneous formation of complex memristic[84] and neuromorphic[85] responses in NP and nanowire assemblies (**Figure 7D**).

To summarize, the study of pathways to complexity in the context of nanomaterials is still at its infancy. Predictions of complexity theories outline an exciting future where complex structures lead to functionality and adaptability. Tools to synthesize such systems, fundamentally understand their formation mechanisms, and predict their functionalities, will be of great value.

**5. Conclusions.** The notion of "complexity" is at the crossroad between the philosophical methodologies of deduction and induction. Deduction has prevailed in the methodology of physics research since the 19th century, where deterministic laws govern a wide variety of phenomena in nature. In contrast, induction is dominant for studies of "emergence" that has gained popularity since the second half of the 20th century, with discoveries of critical phenomena and nonlinear dynamics. Emergence encompasses stochastic behaviors due to many degrees of freedom, the interactions of which are difficult to deduce because they are less deterministic, and the system is non-ergodic.

The concept of structural complexity is at the center of emergence, where a realm of fascinating phenomena arises even when the system components obey very simple rules. Translating this point to materials design, the interactions between the materials components do not need to be excessively complicated and contain multiple parameters to make complex materials. These interactions do need to be, however, multibody, competitive, and experimentally deducible.

Scientists now have the capabilities, therefore, for fundamental and practical studies of complex materials, especially for nanoscale components, that satisfy many requirements and are experimentally convenient. In the next few years, we can aim at better understanding of control and utilization of order, disorder, and hierarchy for specific property sets establishing the links between complexity and functionality. We expect that dynamic nanostructured systems will be created demonstrating gradual emergence of the hierarchical 3D patterns with direct utility in energy technologies, optoelectronics, membrane science, and biomedical applications.

**Acknowledgements**: The authors are grateful for the support from the National Science Foundation (2243104, Center for Complex Particle Systems (COMPASS)), Office of Naval Research (MURI N00014-20-1-2479), the Department of Defense (HQ00342010033, Newton Award for Transformative Ideas during the COVID-19 Pandemic), and Air Force Office of Scientific Research (FA9550-20-1-0265, N.K.).

On behalf of all authors, the corresponding author states that there is no conflict of interest.